\documentstyle[11pt,paspconf,epsf]{article}

\begin{document}

\title{HST Observations of Black Hole X-Ray Transients}

\author{C.A. Haswell and R.I. Hynes}
\affil{Astronomy Centre, University of Sussex, Brighton,
     BN1 9QJ, U.K.}

\begin{abstract}
{Hubble Space Telescope Observations of Black Hole X-Ray Transients
are discussed in the context of the disk instability outburst model.
We focus on the multiwavelength campaign
following GRO~J1655-40 through the summer 1996 outburst.}
\end{abstract}

\keywords{accretion, stars, black holes }

\section{Introduction}

As soon as the class was discovered, the obvious similarities between 
the Black Hole X-Ray Transients (BHXRTs)
and their white dwarf analogues, dwarf novae (DN), guided
investigations into the mechanisms responsible for the dramatic
outbursts exhibited by the former.
The outbursts in DN have been successfully explained
as the result of temperature-dependent viscosity in the accretion
disk: the Disk Instability Model (DIM) (Cannizzo 1993).
The longer recurrence timescales for BHXRTs and the shapes and durations
of their outburst lightcurves, however, provide a challenge
to the DIM (Lasota 1996).

The DIM makes definite quantitative predictions
for the temperature distribution, and hence the expected broad band
spectrum, throughout the outburst cycle 
({\it e.g.} Cannizzo, Chen, \& Livio 1995).
Accretion disk emission is likely to dominate in the UV,
so one of the primary motivations for spectroscopic observations
of BHXRTs with HST is, therefore, to observe the broad band spectral evolution,
and hence address the question of the driving mechanism for
the transient outbursts.
This paper reviews the UV-optical spectra of BHXRTs obtained with HST,
and describes the consequent deductions about the outburst mechanisms.

\section{Observations of Individual Systems}

\subsection{A0620-00}
The first BHXRT to be observed with HST was A0620-00, 16 years after
the 1975 outburst. McClintock, Horne,
\& Remillard (1995) interpreted the ${\rm 1100 - 4500 \AA}$ HST spectrum
in conjunction
with a quiescent ROSAT observation. After subtracting the
contribution of the K5~V mass donor star, they found an optical-UV
accretion spectrum which could be modeled as a 9000~K blackbody, with an area
of only $\sim 1\%$ of the disk area. The low UV flux emitted by this
accreting black hole was a surprise. 
By analogy with quiescent DN a mass transfer
rate into the outer disk of 
${\rm \dot{M}_{disk} \sim 10^{-10}~M_{\odot}~yr^{-1}}$ was inferred.
Meanwhile, the ROSAT soft X-ray flux
implied a mass transfer rate through the inner disk of only
${\rm \dot{M}_{BH} < 5 \times 10^{-15}~M_{\odot}~yr^{-1}}$.
Qualitatively, therefore, these findings were in agreement
with the DIM, suggesting the accumulation of material in the quiescent outer
disk. The extremely low ${\rm \dot{M}_{BH}}$ seemed improbable, however,
and the authors pointed out that isolated black holes might well accrete
more than this from the ISM!

A new explanation was advanced by Narayan, McClintock, \& Yi (1996),
who postulated that the standard disk model is only applicable to the 
outer flow, and that within ${\rm \sim 3000~R_{Sch}}$ the flow is advective:
{\it i.e.} the viscously-generated thermal energy is carried with the
flow rather than being promptly radiated away.
For  black hole accretors, this advected energy can
be carried through the event horizon.
With this hypothesis, therefore, the extremely low quiescent
accretion fluxes do not necessarily demand the extremely low
mass transfer rates inferred from the standard accretion disk model.

\subsection{X-Ray Nova Muscae 1991}

This object was the first to be monitored in the UV-optical through
the decline from outburst, though HST observations occurred only at one
epoch, four months after the maximum. The spectral evolution was analyzed
by 
Cheng et al. (1992).
The data appeared consistent with 
steady-state optically thick accretion disks and the
deduced mass transfer rate fell monotically during the decline.
The DIM predicts, however, that the declining mass transfer rate is
accompanied by a cooling wave propagating through the disk as 
successive hot, high viscosity, annuli make the transition
to the cool, low viscosity, state.
The consequent changing temperature distribution
should have produced an observable cooling wave signature at the long
wavelength end  of the
spectrum. The cooling wave was not observed, however, suggesting problems with
the straightforward application of the DIM to BHXRTs. 

\subsection{GRO J1655-40}

GRO~J1655-40 was discovered in 1994 July;
since then it has undergone repeated outbursts
to a similar level and is apparently an atypical BHXRT.
Superluminal radio jets were associated with the 1994 outburst 
(Hjellming, these proceedings).
Following the onset of X-ray activity in April 1996,
HST spectra were obtained on five separate visits from 1996 May~14
to July~22. 
A full description of these observations and
the associated multiwavelength campaign is given
in Hynes et al. (1997).

GRO~J1655-40 is a highly reddened source, so an accurate correction
for interstellar extinction is a prerequisite to any analysis of
the spectrum. The ${\rm 2175\AA}$ feature 
gives a sensitive
measure of the extinction: E(B-V)=$1.2\pm0.1$, 
a value
consistent with  direct estimates of the visual extinction and
with measurements of interstellar absorption lines
(Hynes et al. 1997).

Figure~1 is the 1996 May~14 dereddened UV-optical spectrum.
Though the UV portion of the spectrum is consistent 
with the $\nu ^{1/3}$ power-law predicted by
the steady-state blackbody disk model, the optical
(${\rm \lambda > 2600 \AA}$) spectrum rises to longer wavelengths
in contrast to the predictions of the model.
Ignoring the ${\rm \lambda >  2600 \AA}$ data, a $\nu ^{1/3}$ model
can be fit to the UV data, leading us to deduce 
the mass transfer rate is 
${\rm 1\times 10^{-7}}$\,M${\rm _{\sun}}$\,yr${\rm ^{-1}\leq
 \dot{M}\leq
 7\times 10^{-6}}$\,M${\rm _{\sun}}$\,yr${\rm ^{-1}}$,
where the dominant source of uncertainty arises from 
interstellar extinction.
Taking
a compact object mass of 7\,M$_{\odot}$ and assuming an accretion efficiency
of ${\rm 10 \%}$, the Eddington rate is 
 ${\rm \dot{M}_{\rm Edd}=1.6\times 10^{-7}}$\,M${\rm_{\sun}}$\,yr${\rm ^{-1}}$,
so near the peak of the outburst this interpretation of the UV spectrum 
implies 
${\rm \dot{M} \approx \dot{M}_{\rm Edd}}$.
\begin{figure}
\label{v1}
\plotone{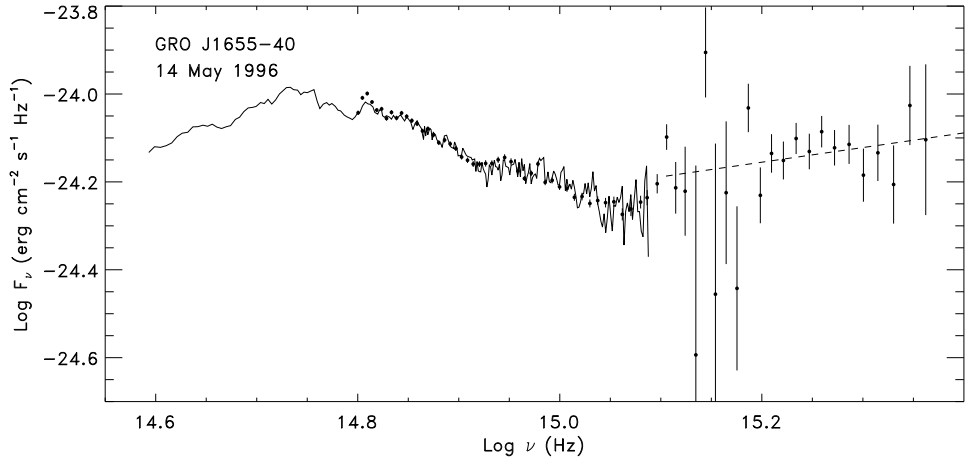}
\caption{The dereddened spectrum of GRO~J1655-40. The spectral slope
changes dramatically at ${\rm log ~ \nu \sim 15.05}$ 
(${\rm \lambda \sim 2600 \AA}$).
The dashed line shows a ${\rm f_\nu \propto \nu ^{1/3}}$
fit to the UV data.}
\end{figure}

We need to invoke something other than a pure steady-state optically thick
accretion disk in order to explain the optical light. The shape
of the spectrum is qualitatively suggestive of an irradiated disk; irradiation
can alter the temperature profile of the outer disk producing a rise
in flux towards longer wavelengths as illustrated in Figure~2. 
The multiwavelength lightcurves for the outburst (Hynes et al. 1997, and 
Hynes et al. these proceedings) do not, however, appear to
support a simple irradiation model: the optical and UV flux declines
while the X-ray flux rises!

\begin{figure}
\label{mwplot}
\plotone{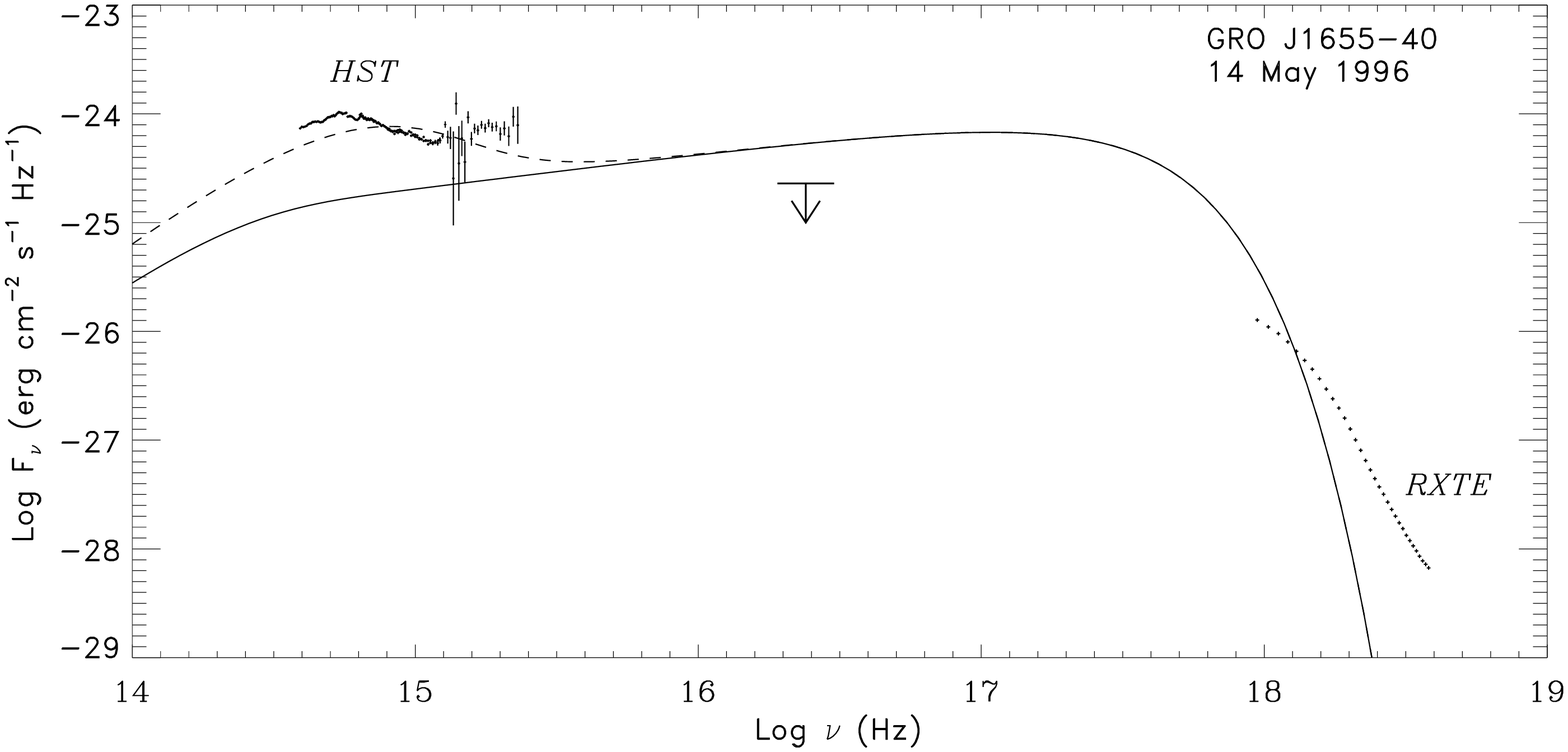}
\caption{Composite spectrum for the May 14 observations. The solid
lines shows a steady state accretion disk spectrum, whilst the dashed line
shows an irradiated disk spectrum. These spectra are
{\bf not} fitted to the data, and are merely illustrative.}
\end{figure}

In order to characterize the optical component of the spectra
we fit blackbody spectra to the ${\rm \lambda > 2600 \AA}$ data for
each visit (Figure~3). While the fluxes fell by about a factor of three between
our first and last visit, the color temperature remained almost
constant, dropping 
from 9800~K to 8700~K; the emitting area dropped from 
${\rm 5.0 \times 10^{23}~cm^2}$
to ${\rm 2.2 \times 10^{23}~cm^2}$.  The system parameters for GRO~J1655-40
are well constrained (Orosz and Bailyn 1997, hereafter OB97, Hjellming and 
Rupen 1995)
and the total available emitting area of the disk and secondary star
is ${\rm \sim 5 \times 10^{23}~cm^2}$, so it is possible to explain
the optical emission at the peak of the outburst as thermal emission,
but only if both the secondary star and the majority of the disk area have
essentially the same isothermal temperature distribution.
Attributing the $\nu ^{1/3}$ UV component to a steady-state disk
is not necessarily ruled out, as this requires only the
hot inner regions of the disk.

\begin{figure}
\plotone{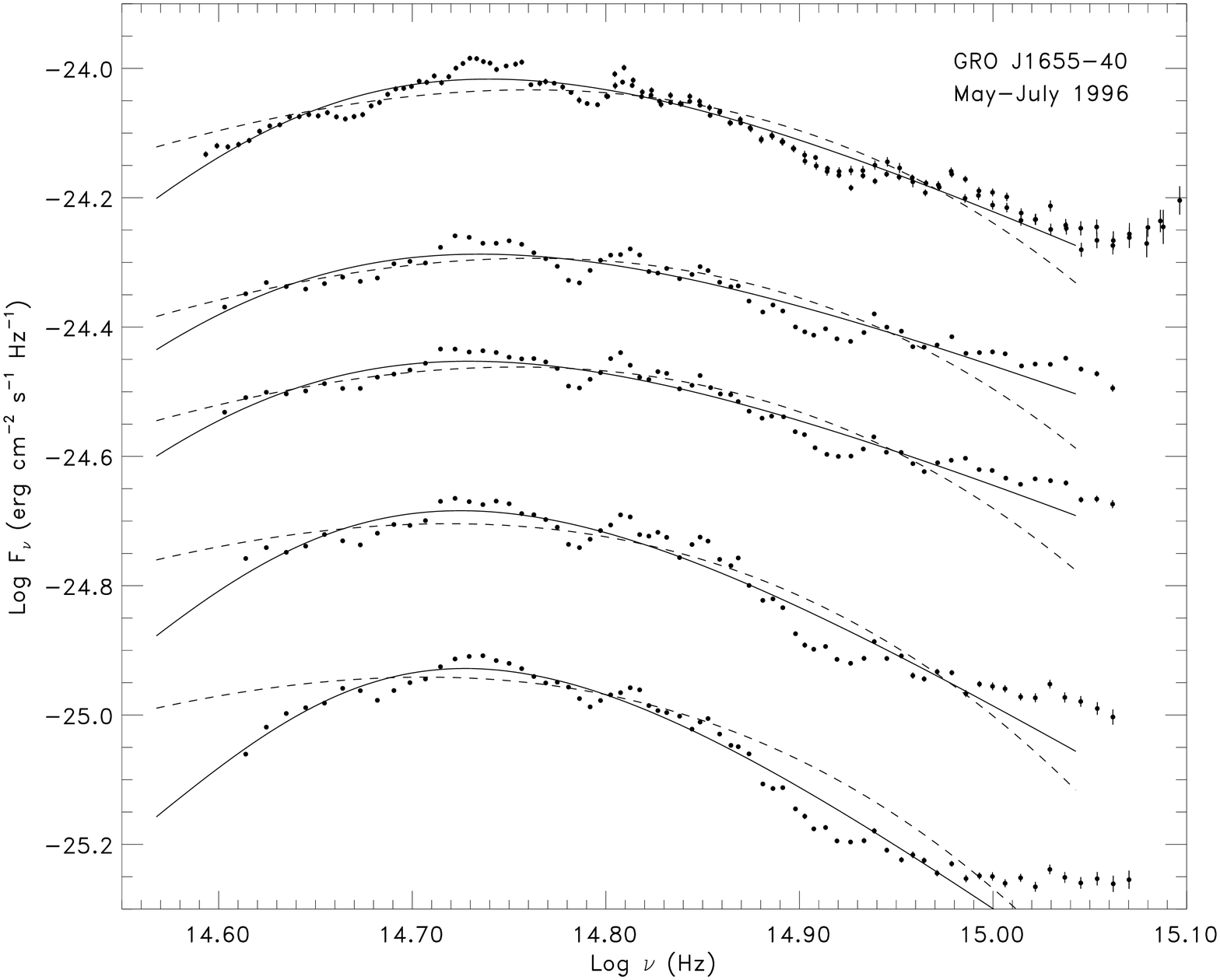}
\caption{Blackbody (dashed line) and synchrotron (solid line) fits to the 
${\rm \lambda > 2600 \AA}$ spectra.
The bumps in the observed spectra are attributable to
spectral line features in the source and to
diffuse interstellar bands.
In order to separate the
         successive visits clearly, a downward shift of 0.1 has been
         introduced in each
         visit relative to the one above it
{\it i.e.}\ the lowest visit has been
         shifted downwards by 0.4.
}
\end{figure}
In addition to the large isothermal area,
there is a suggestion 
that our optical spectra are
more strongly peaked than a single temperature blackbody,
so we considered non-thermal mechanisms. 
Self-absorbed
synchrotron emission from a cloud of relativistic electrons
produced good fits to the optical component; 
the deduced electron energies are ${\rm \gamma \sim 60-90}$, the
magnetic field is ${\rm B \sim 40 - 60~kG}$, and the size
of the cloud is ${\rm 50 - 100~R_{Sch}}$.
Figure~3 shows both the blackbody and the
synchrotron models.
While the synchrotron models fit better, they have an extra free parameter.

Attributing substantial optical 
flux to this compact nonthermal source relieves the requirement
for a large isothermal emitter in the system. It is interesting to
note that intrinsic VRI band linear polarization  ($> 3 \%$)
was detected in July 1996 (Scaltriti et al. 1997), consistent
with the hypothesis of optical synchrotron emission.

On the other hand, the DIM may {\it require} a large isothermal
outer disk for a long-period system like GRO~J1655-40.
Since the disk in such a system is large, the temperature
for a steady state disk falls below
the minimum temperature for the hot, high
viscosity, state long before the outer disk is reached.
Even for an Eddington mass transfer rate in GRO J1655-40
this limit is reached for a radius less than a quarter of
the Roche lobe radius.
This means that there is no global steady-state solution for
${\rm \dot{M} \leq \dot{M}_{Edd}}$. It is not clear what will
happen to the temperature distribution in such 
a case, but it is possible that in outburst much of the outer
disk could be maintained just 
in the hot state. Hence one might expect the outer disk to
appear as an approximately
constant temperature, shrinking area emitter as the decline
proceeds. This hypothesis need to be tested with self-consistent
numerical modeling, but until this is done our data cannot
rule out a thermal interpretation.

The C~IV~${\rm 1550 \AA}$, Si~IV~${\rm 1400 \AA}$, and Si~III~${\rm 1300 \AA}$
UV resonance lines shown in Figure~4 all show likely P-Cygni profiles.
The peak to trough separation in all three cases is 
${\rm \sim 5000 km~s^{-1}}$,
slightly larger than seen in outbursting dwarf novae.
Line profiles produced by
biconical accretion disk winds were
calculated by Shlosman and Vitello (1993) who found `classical' P~Cygni 
profiles
only for inclinations around 60--70$^{\circ}$, in 
striking agreement with the inclination
determined for GRO~J1655-40 (OB97).
We conclude that there was likely a biconical accretion disk wind
at the peak of the UV outburst, when ${\rm \dot{M} \sim \dot{M}_{Edd}}$.
\begin{figure}
\label{pcyg}
\plotone{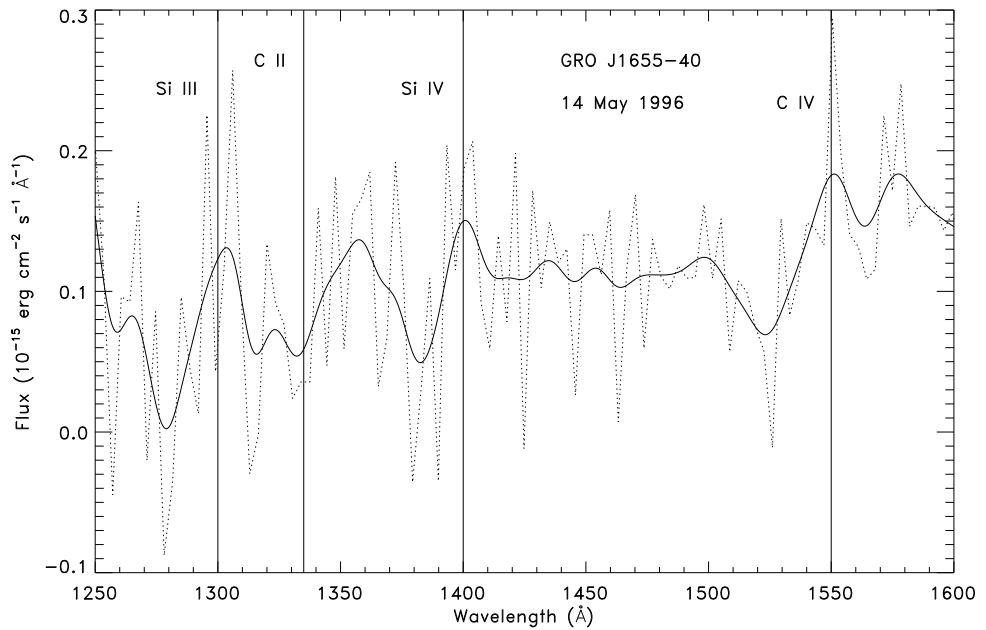}
\caption{Likely P-Cygni profiles detected in the 
UV resonance lines}
\end{figure}


\begin{figure}
\plotone{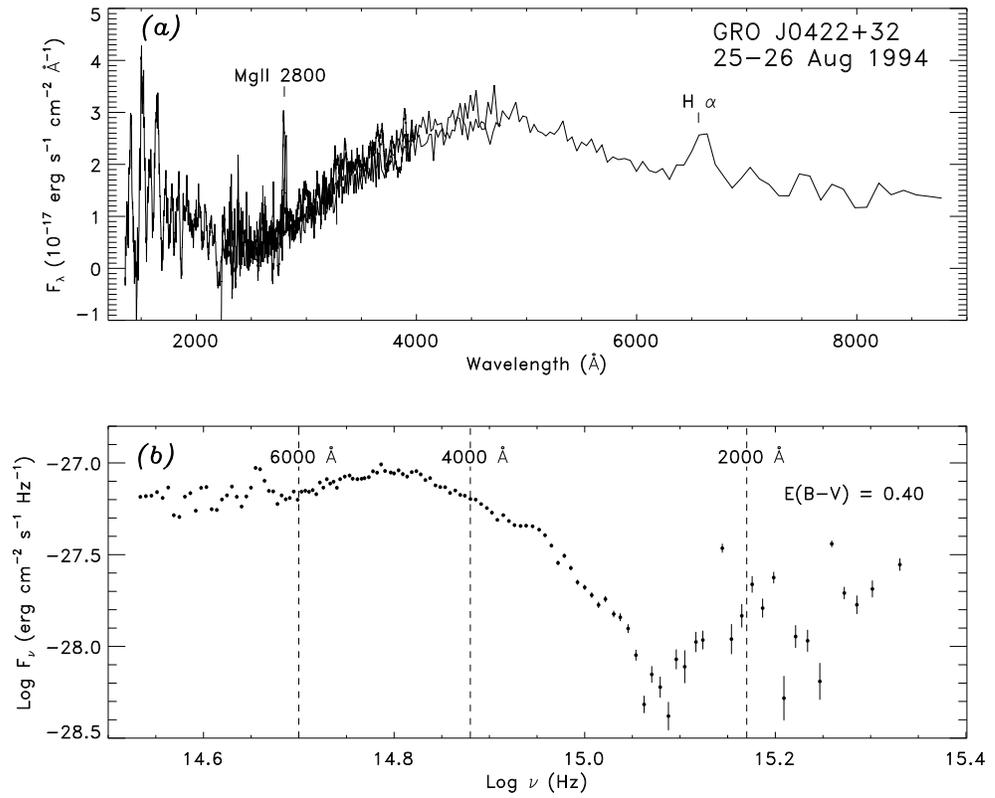}
\caption{(a) The UV-optical spectrum of GRO~J0422+32 in early quiescence.
Much of the light at ${\rm 8000 \AA}$ is due to the mass donor star, but we are
clearly detecting accretion flux at shorter wavelengths.
(b)
The spectrum dereddened using E(B-V)=0.4.
}
\end{figure}
\subsection{GRO J0422+32}
Observations of this target were obtained in early quiescence
approximately two years after the first observed outburst, and seven months
after the last reported reflare (Callanan 1995).
A total of 4.6~hours exposure time was spent obtaining the optical-UV
spectrum with five overlapping spectrograph set-ups.
The target was extremely faint (${\rm R \sim 21}$), so extreme care was 
taken with
the data reduction. The spectrum is shown
in Figure 5(a). 
We note that even if the entire ${\rm 8000 \AA}$ flux is attributed
to the M2~V mass donor (Fillipenko, Matheson, and Ho 1995), the flux at 
${\rm 6000 \AA}$ must be
at least ${\rm 50\%}$ due to the accretion flow, and shortwards
of ${\rm 5000 \AA}$ the accretion light overwhelmingly dominates.

Shrader et al. (1994) obtained multiwavelength data in outburst, 
and noted that the optical-UV energy distribution
showed a pronounced break in slope at ${\rm 4000 \AA}$. 
E(B-V)=0.4 was deduced from the outburst IUE data. 
The quiescent 
HST data, dereddened using  
this value for the interstellar extinction, 
is shown in Figure 5(b).

The spectral shape in Fig.~5(b) is unusual. There is 
a very pronounced change in slope at 
${\rm \sim 2600 \AA}$ 
and the shape
bears a strong resemblance to that
of GRO~J1655-40  in outburst (Figure 1.)
By analogy with the synchrotron fits to the GRO~J1655-40 spectra,
we 
speculate that the strongly rising
optical accretion flux may be partly due to optical synchrotron as
predicted by advection models ({\it e.g.} Narayan, Barret, \& McClintock 1997).

\section{Discussion and Future Work}

The observations described herein have provided
some intriguing challenges to the theoretical models
for transient outbursts. 
As we collect detailed multi-epoch data on more systems
it is becoming clear that the BHXRTs exhibit 
complex and diverse behavior, and it seems 
we need to consider a variety of mechanisms
in order to properly understand the wealth
of observational phenomena.

\acknowledgments

Support for this work
was provided by NASA through grant numbers
GO-6017.01-94A and GO-4377.02-92A
from the Space Telescope Science Institute, which is operated by 
AURA under NASA contract NAS5-26555.
We are grateful to our collaborators, coauthors on Hynes et al. 1997,
for their contributions to the work described herein.

\end{document}